\documentclass[aps, prd, twocolumn, nofootinbib, floatfix]{revtex4}
\usepackage{subfigure}
\usepackage{graphicx}
\usepackage{dcolumn}
\usepackage{epsfig}
\usepackage{amsmath}
\usepackage{amsfonts}
\usepackage{amssymb}
\usepackage{color}
\usepackage{hyperref}
\newcommand{\be}{\begin{equation}}
\newcommand{\ee}{\end{equation}}
\newcommand{\bea}{\begin{eqnarray}}
\newcommand{\eea}{\end{eqnarray}}
\newcommand{\nn}{\nonumber}

\begin{document}

\title{The smearing scale in Laguerre reconstructions of the correlation function}

\author{Farnik Nikakhtar${}^1$}
\email{farnik@sas.upenn.edu}
\author{Ravi K.~Sheth${}^{1,2}$}
\author{Idit Zehavi${}^3$}

\affiliation{${}^1$Department of Physics and Astronomy, University of Pennsylvania, 209 S. 33rd St., Philadelphia, PA 19104 -- USA}
\affiliation{${}^2$Center for Particle Cosmology, University of Pennsylvania, 209 S. 33rd St., Philadelphia, PA 19104 -- USA}
\affiliation{${}^3$Department of Physics, Case Western Reserve University, Cleveland, OH 44106-7079 -- USA}

\date{\today}

\begin{abstract}
  To a good approximation, on large cosmological scales the evolved two-point correlation function of biased tracers is related to the initial one by a convolution.  For Gaussian initial conditions, the smearing kernel is Gaussian, so if the initial correlation function is parametrized using simple polynomials then the evolved correlation function is a sum of generalized Laguerre functions of half-integer order.  This motivates an analytic `Laguerre reconstruction' algorithm which previous work has shown is fast and accurate.  This reconstruction requires as input the width of the smearing kernel.  We show that the method can be extended to estimate the width of the smearing kernel from the same dataset.  This estimate, and associated uncertainties, can then be used to marginalize over the distribution of reconstructed shapes, and hence provide error estimates on the value of the distance scale which are not tied to a particular cosmological model.
  We also show that if, instead, we parametrize the evolved correlation function using simple polynomials, then the initial one is a sum of Hermite polynomials, again enabling fast and accurate deconvolution.  
  If one is willing to use constraints on the smearing scale from other datasets, then  marginalizing over its value is simpler for Hermite reconstruction, potentially providing further speed-up in cosmological analyses.
\end{abstract}


\maketitle

\newcommand{\ste}[1]{\textcolor{red}{\textbf{\small[Ste: #1]}}}


\section{Introduction}\label{intro}
On large ($\sim 100h^{-1}$Mpc) cosmological scales, the evolved two-point correlation function, even of unbiased tracers of the density field, differs in shape from the unbiased linear theory correlation function \cite{rpt, bkPeaks}.  To leading order, this change in shape is due to a (three-dimensional) convolution \cite{Bharadwaj1996}.  For Gaussian initial conditions, the convolution kernel is very well approximated by a Gaussian.  Convolving a polynomial of order $n$ with a Gaussian yields a generalized Laguerre function of order $n/2$, and \cite{LPlaguerre} used this to motivate an algorithm for estimating the initial shape from measurements of the evolved one.  Essentially, fitting a series of half-integer Laguerre functions to the two-point correlation function of biased tracers of the cosmological density field allows one to perform the deconvolution analytically, even in the presence of redshift space distortions \cite{HODlaguerre}. 

This reconstruction of the initial shape requires as input a guess for the width of the Gaussian smearing kernel $\Sigma$.  In CDM models, the physics which gives rise to the convolution relates the smearing scale to the background cosmological model:
\begin{equation}
 \Sigma^2 = \int dk\, P_{\rm L}(k,z)/3\pi^2,
 \label{smear}
\end{equation}
where $P_{\rm L}(k,z)$ is the linear theory power spectrum \cite{rpt} at redshift $z$. So, if the model parameters are sufficiently well-known, then they can be used to provide a useful estimate of the smearing scale.  However, it is interesting to ask if the data constrain this scale independently -- i.e., without having to assume a fiducial cosmological model.  This is particularly interesting because the uncertainties on this estimate can be propagated into uncertainties on the shape of the reconstructed correlation function.  In turn, this provides a means to marginalize over the value of the smearing scale when estimating cosmological parameters -- such as the cosmological distance scale at the redshift of the survey -- that is not tied to a fiducial model.

In Section~\ref{sec:data2smear}, we show that this is indeed possible in principle, but whether or not the constraint which results is sufficiently tight to be interesting (better than $\sim 10$\% precision) will depend on the data set (volume and tracer number density).  We illustrate the method using the real space dark matter correlation function, before applying it to the monopole of the redshift space distorted correlation function measured in mock galaxy catalogs.  

Section~\ref{sec:app} shows how this fiducial model-free determination of the smearing scale can be used to estimate the cosmological distance scale and, in particular, to provide realistic estimates of the precision of this determination.  If the smearing scale is not well-constrained, then this decreases the precision of the distance scale constraint.  Therefore, especially for small surveys, it may be that other datasets provide more useful determinations of the smearing.  We argue that this motivates consideration of other parametrizations of the reconstruction problem.  E.g., even if useful constraints on the smearing scale must come from independent datasets, some parametrizations of the reconstruction problem may allow one to marginalize over the value of the smearing scale more easily than others. These alternative parametrizations are the subject of Section~\ref{sec:hermites}.  A final section summarizes our results.  

We illustrate our arguments using the dark matter correlation functions measured in the $z=1$ outputs of the Quijote simulations \cite{Quijote_sims}, as well as mock galaxy catalogs in the $z=0$ outputs of these same simulations.  Each of these 15000 simulations followed the evolution of $512^3$ particles in a periodic box of side $L=1h^{-1}$Gpc (comoving).  The fiducial cosmology of this set is flat $\Lambda$CDM with 
$(\Omega_m,\Omega_b, h, n_s, \sigma_8) = (0.3175, 0.049, 0.6711, 0.9624, 0.834)$, for which $\Sigma$ of equation~(\ref{smear}) equals $5.1h^{-1}$Mpc at $z=1$ and $8.5h^{-1}$Mpc at $z=0$. 
To explore how our results scale with effective volume, we average together the correlation functions measured in 10, 50, and 100 boxes at a time to crudely mimic 1500, 300, and 150 realizations of effective volumes of 10, 50, and 100~($h^{-1}$Gpc)$^3$ each.
In all cases, the correlation functions in these boxes are measured in bins of width $1h^{-1}$Mpc, and all our analyses only make use of the scales between $70$ and $110h^{-1}$Mpc.  On these scales, the covariance between the different scales is well described by a simple `smeared linear theory plus Poisson shot-noise model' \cite{eppur, LPnus, LPlaguerre}.

\section{Smearing scale from data}\label{sec:data2smear}
Our starting point is that the evolved pair correlation function is related to that predicted by linear theory (i.e. the initial one multiplied by a growth factor) by a convolution:
\begin{equation}
 \xi_{\rm NL}(\mathbf{s}) = \xi_{\rm L}\otimes G + \xi_{\rm MC}(\mathbf{s})
    \approx \int d\mathbf{r}\,\xi_{\rm L}(\mathbf{r})\,G(\mathbf{s-r}|\Sigma),
 \label{xiRPT}
\end{equation}
e.g., \cite{rpt}, where the final expression assumes that the `mode-coupling term' $\xi_{\rm MC}$ can be ignored.  We will ignore $\xi_{\rm MC}$ in all of the analysis which follows, since none of the main points we make are changed if we include it.

\subsection{`Optimal' estimate}
We begin with an exploration of the precision with which the smearing scale can be estimated if the shape and amplitude of the linear theory power spectrum are known.  This means that we simply fit the right hand side of equation~(\ref{xiRPT}) to the measured $\xi_{\rm NL}$ to determine the value of $\Sigma$.  The fitting uses measurements in bins of width $1h^{-1}$Mpc over the range $70$-$110h^{-1}$Mpc, and uses the covariance matrix described in \cite{LPlaguerre} to account for the fact that bins in $\xi_{\rm NL}$ are correlated.  

\begin{figure}
 \centering
 \includegraphics[width=1\hsize]{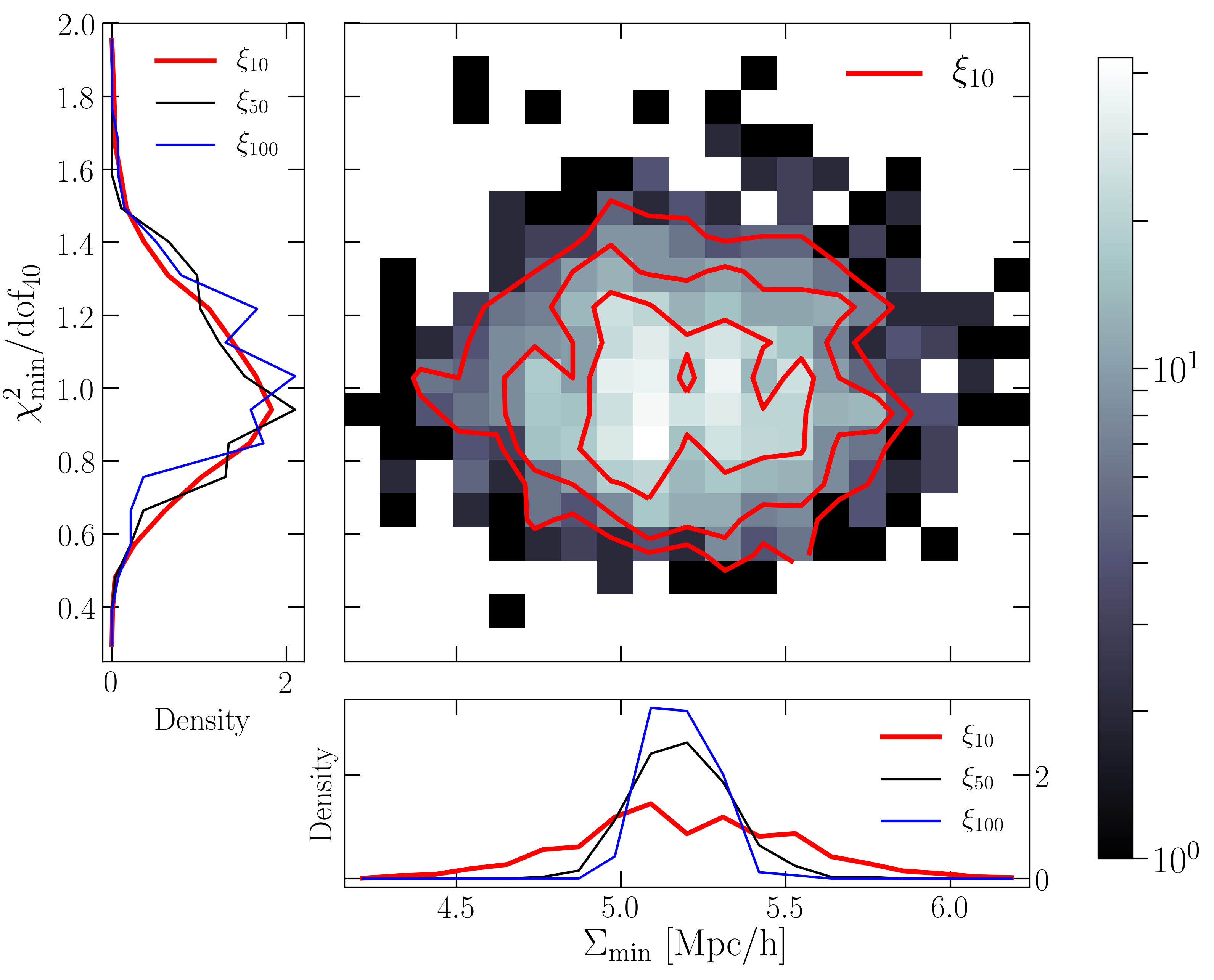}
 \caption{\label{fig:pdfSmear} 
   Estimated $\Sigma$ when we use exactly the correct shape and amplitude of $\xi_{\rm L}$ when fitting the right hand side of equation~(\ref{xiRPT}) to the measured $\xi_{\rm NL}$ but $\Sigma$ is determined from the fit, as well as a measure of the goodness of fit, $\chi^2_{\rm min}$/dof.  Main panel shows this joint distribution from 1500 realizations of a volume of size $10h^{-1}$Mpc.  Histograms in bottom panel show the distribution of $\Sigma$ for three different effective volumes as indicated (10, 50 and $100h^{-1}$Mpc):  larger volumes return a narrower distribution, hence a tighter constraint on the smearing scale.  Histograms in the panel on the left show the associated distributions of $\chi^2_{\rm min}$/dof.}
\end{figure}

The two-dimensional histogram in Figure~\ref{fig:pdfSmear} shows the joint distribution of the best-fitting $\Sigma$ and the associated value of $\chi^2_{\rm min}$/dof constructed from the 1500 realizations of an effective volume of $10h^{-3}$Gpc$^{3}$.  The red curves in the left hand panel show the distribution of $\chi^2_{\rm min}$/dof values obtained by projecting out the $\Sigma$ values.  It peaks close to unity, indicating that the fits are generally acceptable.  The red curves in the bottom panel show the distribution of best-fit $\Sigma$ values, obtained by projecting out the $\chi^2_{\rm min}$ values.  It peaks at about $5.2h^{-1}$Mpc, which is slightly higher than the linear theory value of $5.1h^{-1}$Mpc, consistent with previous work \cite{rpt}.  The rms of this distribution is $0.33h^{-1}$Mpc, which is about a 7\% precision estimate of the smearing scale.  The other curves show similar projections of the histograms made from 300 and 150 realizations of effective volumes of size $50$ and $100h^{-3}$Gpc$^{3}$.  Clearly, larger volumes are more constraining:  the rms values are 0.14 and $0.1h^{-1}$Mpc, scaling approximately with the inverse of the square root of the effective volume (Appendix~\ref{sec:chi2} discusses why).  Thus, Figure~\ref{fig:pdfSmear} suggests that, if the shape and amplitude of $\xi_{\rm L}$ are known, then the expected precision on the estimated $\Sigma$ is about $3\, \sqrt{50h^{-3}{\rm Gpc}^3/V_{\rm eff}}$ percent.  Allowing for the amplitude of the power spectrum to be a free parameter when fitting only slightly degrades the precision of the $\Sigma$ estimate. (There is a slight degeneracy between the amplitude and the smearing scale -- a higher initial peak must be smeared more to produce the same observed amplitude -- but because the smearing matters little at scales $\sim 60h^{-1}$Mpc, the degeneracy is reduced by including scales that are far from the peak when fitting.)

That said, in practice, the biased tracers will be less abundant than the dark matter, so although the shape of the correlation function of biased tracers should not be too different from that of the dark matter, the measurement errors will be larger.  This will degrade the precision of the constraint on $\Sigma$, which will propagate to other analyses which use its value. For example, in the context of reconstructing the shape of $\xi_{\rm L}$, the distribution shown in Figure~\ref{fig:pdfSmear} could be used as a prior on the value of the smearing scale when averaging over distributions such as those shown in Figure~8 of Ref.\cite{LPlaguerre}.  Nevertheless, we think further analysis of this estimator of $\Sigma$ is potentially interesting.  In particular, the next subsection explores the accuracy and precision of estimated $\Sigma$ if neither the shape nor the amplitude of $\xi_{\rm L}$ are known a priori.  


\subsection{Laguerre-based estimate}
If $\xi_{\rm L}$ is a function of $|\mathbf{r}|$ and the three-dimensional Gaussian smearing kernel is isotropic with rms $\Sigma$ in each direction, then 
\begin{equation}
  \xi_{\rm NL}(s) 
  = \int_0^\infty \frac{dr\,r^2}{\Sigma^3}\,\frac{{\rm e}^{-(r^2+s^2)/(2\Sigma^2)}}{\sqrt{2\pi}}\,2\frac{\sinh(rs/\Sigma^2)}{rs/\Sigma^2}\,\xi_{\rm L}(r).
 \label{xiNLconv}
\end{equation}
The terms other than $\xi_{\rm L}$ in the integral define a noncentral-Chi distribution in $r/\Sigma$ with 3 degrees of freedom and noncentrality parameter $s/\Sigma$.  So, if $\xi_{\rm L}$ is written as a sum of polynomials, then $\xi_{\rm NL}$ is a sum over moments of the $\chi_3$ distribution -- generalized Laguerre functions.  I.e., if 
\begin{equation}
  \xi_{\rm L}(r) = \sum_{k=0}^n a_k\, (r/{\cal R})^k
  \label{xiLin}
\end{equation}
for some set of coefficients $a_k$ (${\cal R}$ merely serves to make the $a_k$ dimensionless -- it plays no fundamental role), then
\begin{equation}
  \xi_{\rm NL}(s) = \sum_{k=0}^n a_k\, (s/{\cal R})^k\,(\Sigma/s)^k\,\mu_k(s/\Sigma),
  \label{xiLag}
\end{equation}
where
\begin{align}
  \mu_{2n} &= 2n!!\, L_{n}^{(1/2)}(-x^2/2) \nn \\
  \mu_{2n-1} &= (2n-1)!!\,\sqrt{\frac{\pi}{2}}\,L_{n-1/2}^{(1/2)}(-x^2/2),
 \label{chi3-moments}
\end{align}
and the $L_\beta^{(\alpha)}(x)$ are generalized Laguerre functions.  This shows that the shape of $\xi_{\rm NL}$ differs from $\xi_{\rm L}$ because $\mu_k(x)/x^k \ne 1$.  

Ref.\cite{LPlaguerre} made the point that, if the $a_k$ are determined by fitting equation~(\ref{xiLag}) to the observed $\xi_{\rm NL}(s)$, then the `Laguerre reconstructed/deconvolved' shape $\xi_{\rm Lag}(r)$, is given by inserting the fitted $a_k$ in equation~(\ref{xiLin}).  If $\Sigma$ is known, then the fitting reduces to a simple linear least squares problem.  In equation~(\ref{xiLag}), the terms which involve $\Sigma$ multiply the $a_k$, so if $\Sigma$ must also be determined from the fitting process, then the problem to be solved is nonlinear, but there is no other complication.  

To illustrate, we have fit equation~(\ref{xiLag}), with $n=9$, to the same $\xi_{\rm NL}$ measurements we show in Figure~\ref{fig:pdfSmear} for a number if choices of $\Sigma$:  i.e. for each $\Sigma$, we solve a linear least squares problem.  We then compare the $\chi^2_{\rm min}$ values and choose that $\Sigma$ for which $\chi^2_{\rm min}$ is smallest.  (We have checked that the values of $\chi^2_{\rm min}$/dof are consistent with unity, indicating the fits are acceptable.)  The histograms show the distribution of estimated $\Sigma$ values
for effective volumes of 10, 50, and 100 ~($h^{-1}$Gpc)$^3$ (red, black, and blue, respectively).  In all three cases, the mean values are slightly larger than the linear theory value of $5.1h^{-1}$Mpc, by about the same amount as in the bottom panel of Figure~\ref{fig:pdfSmear}.  The rms values of the distributions are 0.46, 0.21 and $0.15h^{-1}$Mpc: They are about 50 percent larger than in the bottom panel of Figure~\ref{fig:pdfSmear} when the shape and amplitude were fixed to their correct values. This is not surprising:  the Laguerre-based approach must determine the amplitude {\em and} shape of $\xi_{\rm L}$ as well as the value $\Sigma$.  Viewed from this perspective, the Laguerre-based approach does rather well.

\begin{figure}
 \centering
 \includegraphics[width=1\hsize]{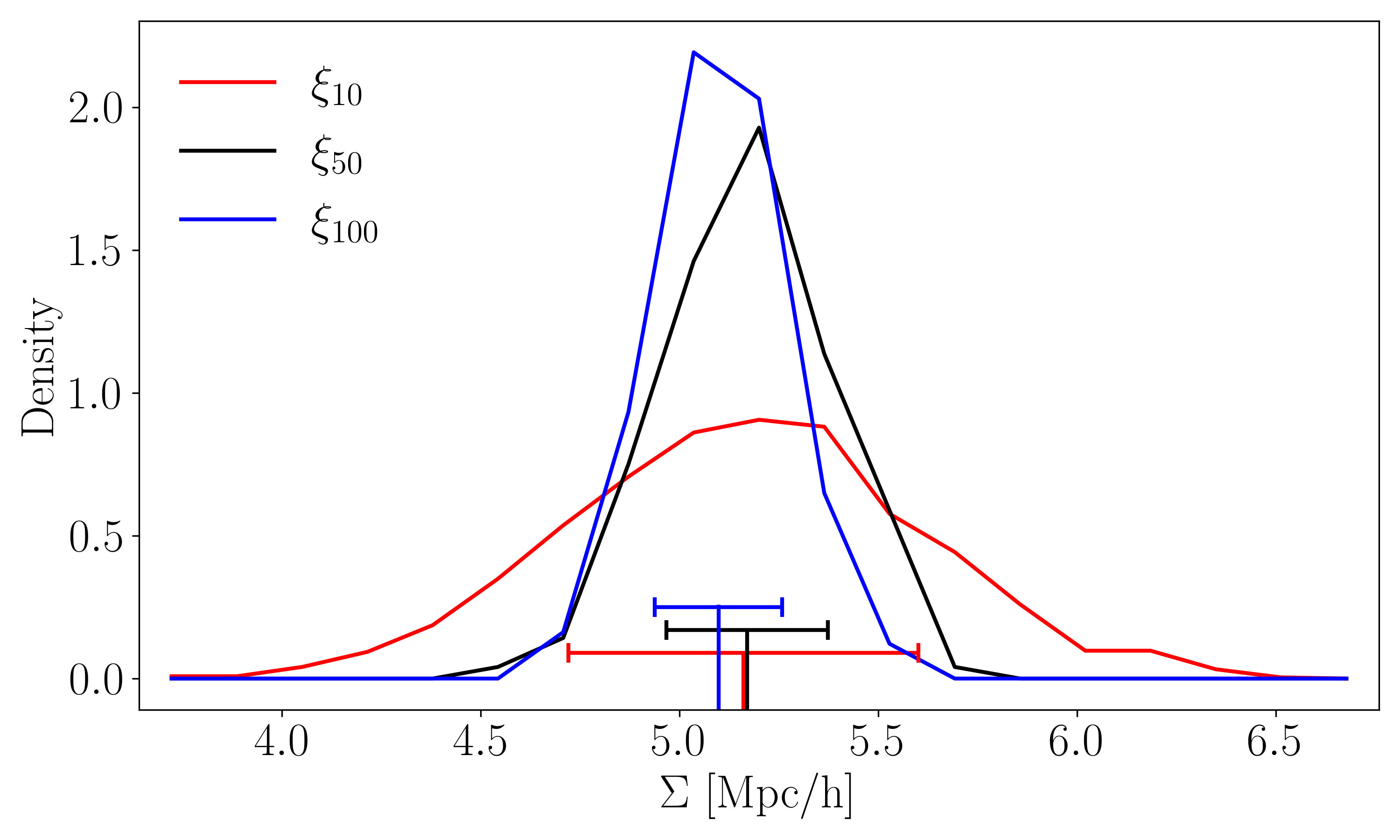}
 \caption{\label{fig:lagSmear} 
   Distribution of estimated $\Sigma$ from our Laguerre-based analysis (fit equation~\ref{xiLag} to the measured $\xi_{\rm NL}$).  Three histograms show results for three different effective volumes as indicated:  larger volume returns a narrower distribution, hence a tighter constraint on the smearing scale.}
\end{figure}

Since the Laguerre-based estimate of $\Sigma$ could be artificially broadened if there are no parameter choices for which equation~(\ref{xiLin}) can provide a good description of the linear theory shape, 
it is interesting to search for other parameterizations of $\xi_{\rm L}$ which may constrain $\Sigma$ better.  We study this in Section~\ref{sec:hermites}.

\subsection{Illustration using biased tracers}
So far, we have focussed on the dark matter correlation function in configuration space.  Real data of rare, biased tracers will include redshift-space distortions and shot-noise.  To study how our constraints degrade in a more realistic setting, we have explored the following extreme scenario:  We work with $\xi_0$, the monopole of the redshift-space distorted correlation function of mock galaxies from the same Quijote simulation set (the Molino suite of mock galaxy catalogs \cite{molino-mocks}), but now at $z=0$. These mock catalogs use the standard \cite{Zheng2007} Halo Occupation Distribution (HOD) model. The number density of the mock galaxies is $1.63 \times 10^{-4}h^{-3}$Mpc$^3$, so the shot-noise is significantly larger than for the dark matter, and the bias factor $b=2.4$.  In addition, the lower redshift means that the smearing scale is larger:  $\Sigma = 8.5h^{-1}$Mpc.  Moreover, the fact that we are working in redshift-space means that the smearing scale and the bias factor are modified to
\begin{equation}
 b^2_{\rm eff} \approx b^2 \left[1 + 2 \beta/3 + \beta^2/5\right] 
 \label{bkaiser}
\end{equation}
and
\begin{equation}
 \Sigma_{\rm eff}^2 = \Sigma^2 \left[1 + \frac{f(2+f)}{3}\frac{1 + 6\beta/5 + 3\beta^2/7}{1 + 2\beta/3 + \beta^2/5} \right] ,
 \label{zmear}
\end{equation}
where $f\equiv d\ln D/d\ln a = 0.53$ and $\beta \equiv f/b = 0.22$ \cite{HODlaguerre}.  In our mocks, $b_{\rm eff}=2.6$ and $\Sigma_{\rm eff}=10.34h^{-1}$Mpc.  

In practice, the Laguerre method assumes that equation~(\ref{xiNLconv}) still applies with $\xi_{\rm L}\to b_{\rm eff}^2\xi_{\rm L}$ and $\Sigma\to\Sigma_{\rm eff}$.  So the question is:  How well does the method recover $\Sigma_{\rm eff}$, while also recovering the shape and amplitude of $b_{\rm eff}^2\xi_{\rm L}$?

\begin{figure}
 \centering
 \includegraphics[width=1\hsize]{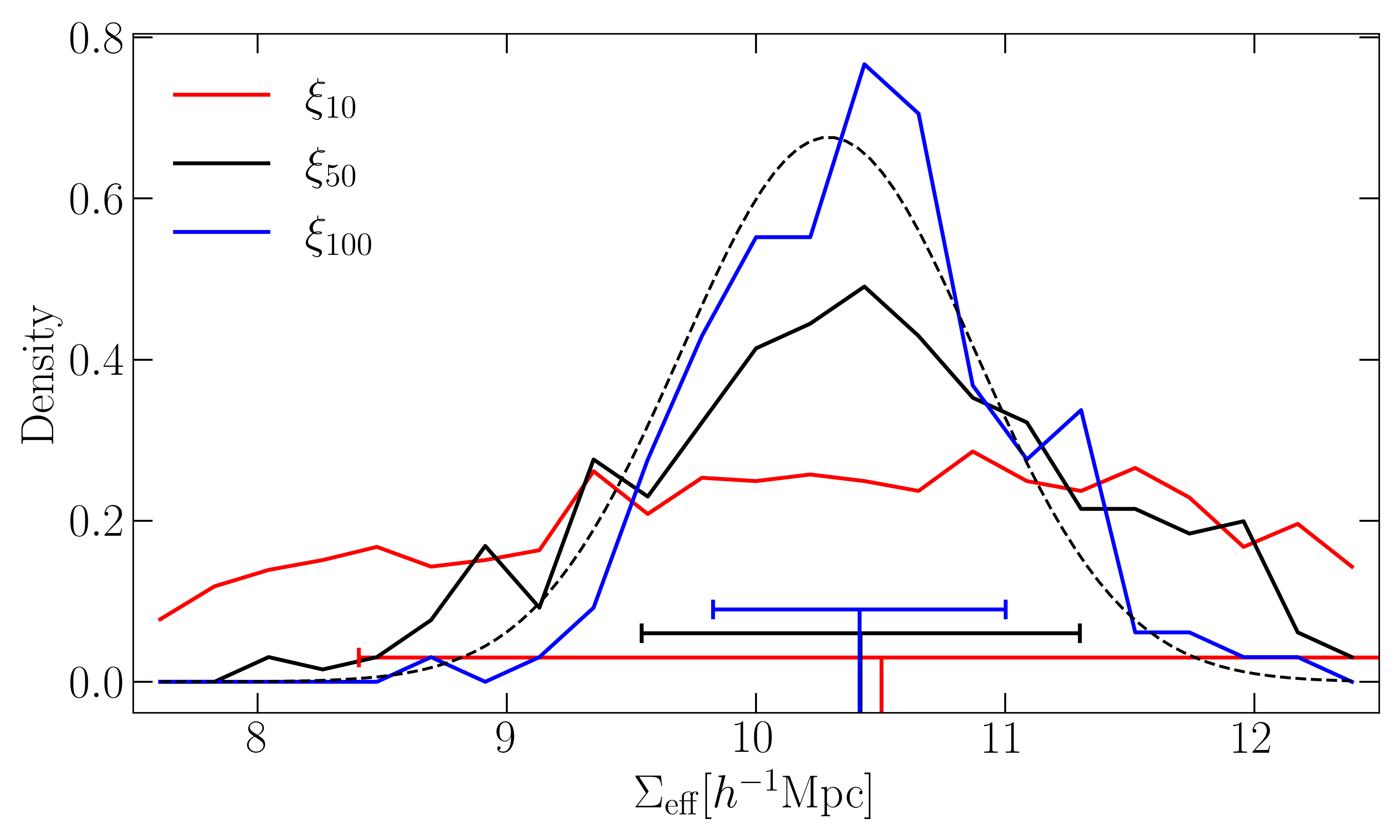}
 \caption{\label{fig:hodZmear} 
   Distribution of estimated $\Sigma_{\rm eff}$ from our Laguerre-based analysis (fit equation~\ref{xiLag} to the measured redshift-space correlation function).  Three histograms show results for 1500, 300 and 100 realizations of the three different effective volumes as indicated (10, 50 and 100$h^{-3}$Gpc$^3$).  Vertical bars near the bottom mark the mean values of the distributions; horizontal error bars show the mean $\pm$ the rms.  Dashed curve shows the constraint on $\Sigma_{\rm eff}$ from propagating the error on a single realization of the $\xi_{50}$ ensemble (the one shown in Figure~\ref{fig:hodrec}).}
\end{figure}

Figure~\ref{fig:hodZmear} shows the result of estimating $\Sigma_{\rm eff}$ by fitting 9th order Laguerres to the 1500, 300 and 100 realizations of $\xi_0$
(effective volumes of 10, 50 and 100($h^{-1}$Gpc)$^3$).  The method returns distributions are that are centered at $\Sigma_{\rm eff}=10.50, 10.41$, and 10.42$h^{-1}$Mpc for the smallest to largest volumes.  These are all close to the theory value of $\Sigma_{\rm eff}=10.34h^{-1}$Mpc. The rms scatter around the mean is about 20, 8.5, and 6\% of the mean value; while this is significantly worse than for the dark matter at $z=1$, it is still a sub-ten percent determination for DESI-like volumes.

Perhaps more importantly, these values for the rms are similar to those returned by the fitting procedure when fitting a single simulation.  Since the distribution of $\Sigma$ shown in Figure~\ref{fig:hodZmear} is approximately Gaussian, one can approximate $p(\Sigma)$ by assuming it is Gaussian and treating the uncertainty on $\Sigma$ determined from a single realization as its rms.  The dashed curve in Figure~\ref{fig:hodZmear} shows an example:  it was obtained by propagating the errors on the fit, shown as a black dashed curve, to the symbols in Figure~\ref{fig:hodrec} (which show one member of the $\xi_{50}$ ensemble).

Because the Laguerre methodology is agnostic about the shape or amplitude of $\xi_{\rm L}$, the $p(\Sigma)$ which it returns -- essentially the shape shown in Figure~\ref{fig:hodZmear} -- is obtained without assuming anything about the linear theory power spectrum.  

\section{Applications}\label{sec:app}
The previous section showed that the Laguerre methodology is able to provide useful estimates of the smearing scale that are not tied to a fiducial cosmology.   We now discuss what additional science such estimates enable.

\subsection{Accuracy and precision of reconstructed distance scale}\label{sec:rec}
\begin{figure}
 \centering
 \includegraphics[width=1\hsize]{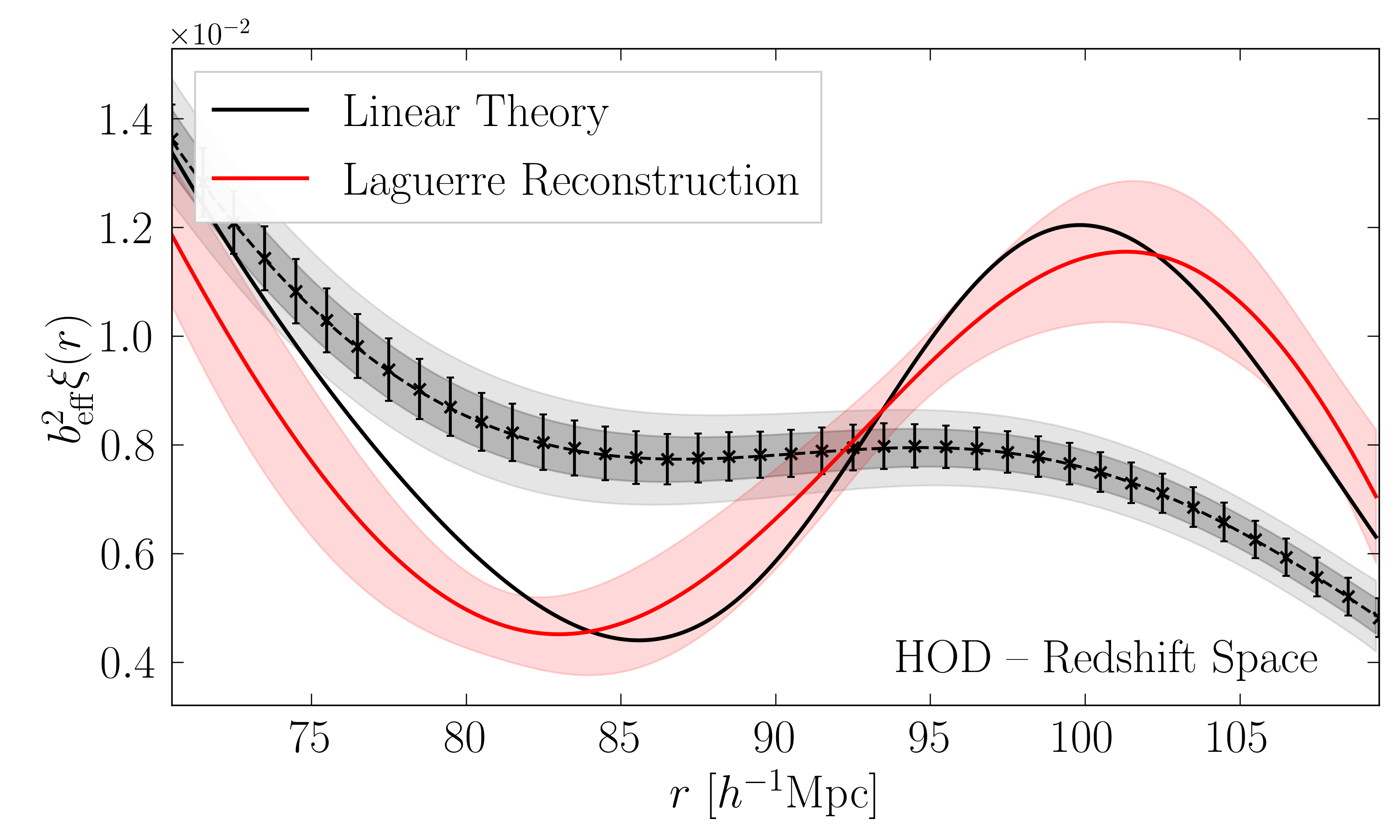}
 \caption{\label{fig:hodrec} 
   Measured redshift-space monopole for a mock galaxy catalog in an effective volume of 50$h^{-1}$Gpc (symbols with error bars), 9th order Laguerre fit to it (black dashed curve, which shows equation~\ref{xiLag}), and associated reconstruction (red dashed curve, which shows equation~\ref{xiLin} with the $a_k$ determined by the black dashed curve).  This particular realization happens to have $\Sigma_{\rm eff}=10.3h^{-1}$Mpc.  Grey bands show the 68\% and 95\% range covered by 300 such realizations of the observed redshift-space monopole, and pink bands show the 68\% range covered by the 300 associated reconstructed shapes $\xi_{\rm Lag}$.  Solid black curve shows the linear theory $b^2_{\rm eff}\xi_{\rm L}$.}   
\end{figure}

Figure~\ref{fig:hodZmear} shows the distribution of $\Sigma_{\rm eff}$ values at which equation~(\ref{xiLag}) best fits the measured $\xi_0$.  However, the best fit also determines a set of coefficients $a_k$ which, when inserted in equation~(\ref{xiLin}), determine the reconstructed shape $\xi_{\rm Lag}$.  To illustrate, the black symbols (with error bars) in Figure~\ref{fig:hodrec} show a single realization of $\xi_0$ in a $50h^{-3}$Gpc$^3$ volume, and the dashed black curve shows the best-fit of equation~(\ref{xiLag}) to it.  This fit determines the coefficients $a_k$ as well as $\Sigma_{\rm eff}$.  This particular realization has $\Sigma_{\rm eff} = 10.3h^{-1}$Mpc.  The grey bands show the regions which enclose 68\% and 95\% of 300 realizations of $\xi_0$ in the same effective volume.  Notice that the measurement (symbols) shows almost no peak or dip because, by $z=0$, particles have moved far from their initial positions, and their speeds (which give rise to redshift space distortions) are also large. This is also true of the ensemble average bracketed by the grey regions; in fact, in about 8\% of the simulations, there is no discernable peak or dip.

Despite this extreme smearing, the Laguerre reconstructed shape $\xi_{\rm Lag}$ (red curve shows equation~\ref{xiLin} with the best fit $a_k$) is reasonably close to that of linear theory $\xi_{\rm L}$ (black solid).\footnote{Note that because we are ignoring the mode-coupling contribution the reconstructed shape is not as good as it could possibly be.  However, because our main concern here is to illustrate the qualitative effects of $\Sigma$, we will continue to ignore mode-coupling.}  The pink bands show the region which encompasses 68\% of the 300 reconstructed $\xi_{\rm Lag}$ curves -- i.e. the reconstructions of the curves which resulted in the grey bands.  

To turn each of these reconstructed shapes into an estimate of the cosmological distance scale (at the survey redshift) we use the `linear point' $r_{\rm LP}$, which is defined as the midpoint between the peak and dip scales in the correlation function:
\begin{equation}
 r_{\rm LP} \equiv \frac{r_{\rm peak} + r_{\rm dip}}{2}.
 \label{eq:rLP}
\end{equation}
For the background cosmology of our mock catalogs, $r_{\rm LP}=92.7h^{-1}$Mpc in linear theory.  Ref.~\cite{PaperI} argue that this scale is interesting because $r_{\rm LP}$ in the evolved correlation function is approximately the same as in $\xi_{\rm L}$.  Distance scale estimates boil down to estimating $r_{\rm LP}$ in each $z=0$ mock and providing a realistic error bar for it.  

Although $r_{\rm LP}$ evolves less than either the peak or dip scales \cite{PaperI,LPlaguerre}, it does shift to slightly smaller scales at later times.  In the $z=0$ mocks we are studying here, the smearing is so large that there is no discernable peak or dip in about 8\% of the simulations having effective volumes of
50$h^{-3}$Gpc$^3$.  (For the 10 and 100$h^{-3}$Gpc$^3$ volumes, this fraction is 9\% and 6\% and they are not used in the reconstruction process.)  But in the others, the values of $r_{\rm LP}$ estimated from the Laguerre fit (i.e. by finding the maximum and minimum of equation~\ref{xiLag} using the best-fitting values of $a_k$ and $\Sigma$) are centered on $90.7h^{-1}$Mpc, with an rms scatter of about $\pm 1.3h^{-1}$Mpc.  This is a significant offset from its linear theory value,  
so the question is if the $r_{\rm LP}$ estimates from reconstructed $\xi_{\rm Lag}$ are closer to the linear theory value of $92.7h^{-1}$Mpc, and what is the associated uncertainty on this `reconstructed' value.

\begin{figure}
 \centering
 \includegraphics[width=1\hsize]{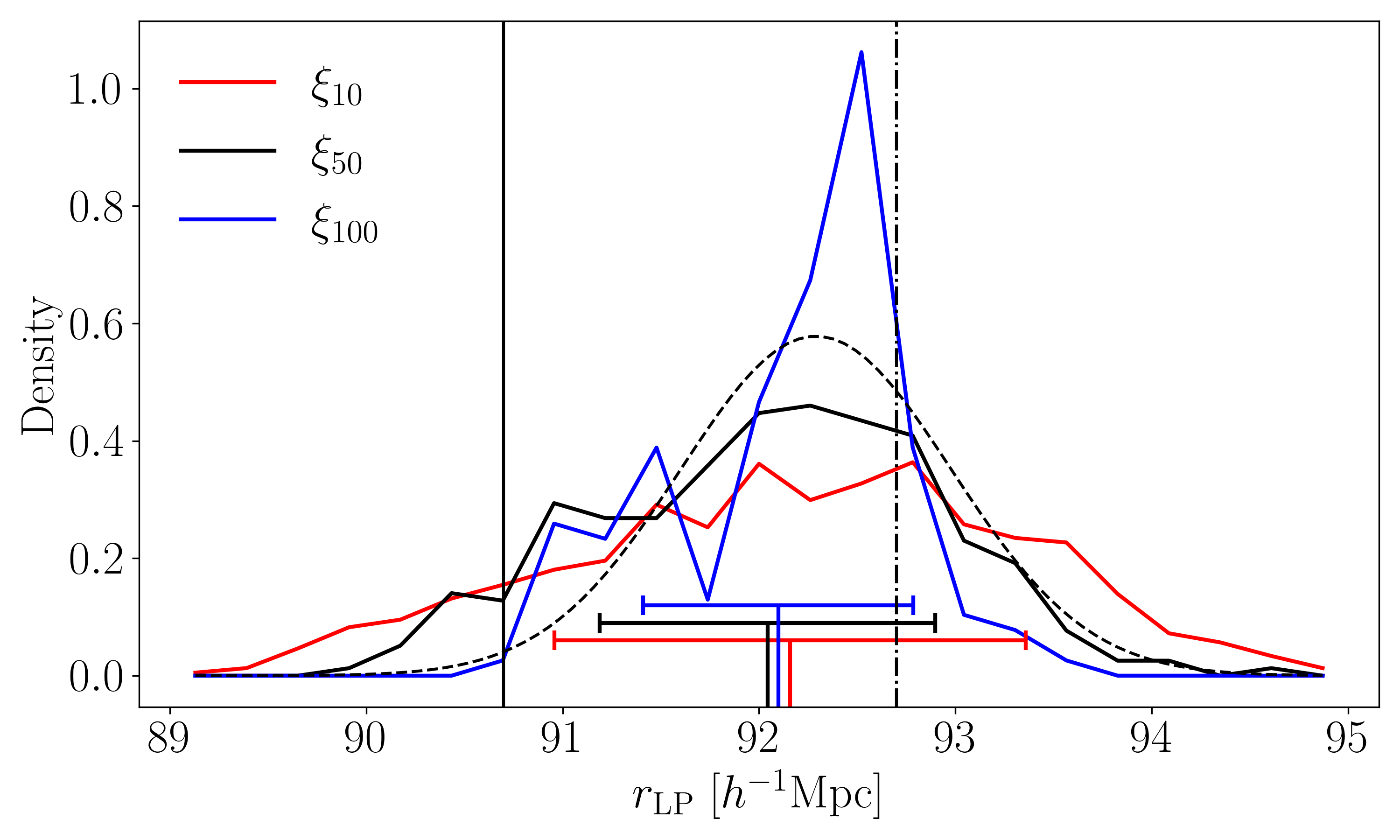}
 \caption{\label{fig:hodrLP} 
   Distribution of reconstructed $r_{\rm LP}$ from our Laguerre-based analysis (fit equation~\ref{xiLag} to $\xi_0$, the monopole of the measured redshift-space correlation function; then insert the $a_k$ coefficients of the best fit in equation~\ref{xiLin} to define $\xi_{\rm Lag}$; finally, determine $r_{\rm LP}$ from the peak and dip scales of $\xi_{\rm Lag}$).  Three histograms show results for three different effective volumes as indicated.  Vertical bars near the bottom mark the mean values of the distributions; horizontal error bars show the mean $\pm$ the rms.  Smooth black dashed curve shows a Gaussian distribution with mean and rms determined from the single $\xi_{\rm Lag}$ shown in Figure~\ref{fig:hodrec}.  Vertical solid line shows the mean $r_{\rm LP}$ scale measured in $\xi_0$ prior to reconstruction (it is similar for all three effective volumes), and dot-dashed line shows the scale in linear theory.}
\end{figure}

The solid black curve in Figure~\ref{fig:hodrLP} shows the distribution of reconstructed $r_{\rm LP}$ values determined from the $\xi_{\rm Lag}$ that were reconstructed from the same $\sim 300$ $\xi_{50}$ measurements which led to Figure~\ref{fig:hodZmear}.  
The black vertical bar near the bottom marks the mean value of the distribution and the horizontal error bar shows the mean $\pm$ the rms.  The smooth dashed curve shows a Gaussian distribution with mean and variance determined from propagating the errors on $r_{\rm LP}$ for a single member of this ensemble -- the one shown by the symbols in Figure~\ref{fig:hodrec}, that is best fit by the black dashed curve there, and whose reconstructed $\xi_{\rm Lag}$ is the dashed red curve there.  (Whereas the mean and variance come from standard error analysis, using a Gaussian shape is an extra assumption; this is reasonable as the distribution defined by the ensemble is not too non-Gaussian.)  The mean of this single realization is consistent with the ensemble mean reconstructed value which is centered on $\sim 92.1h^{-1}$ with an rms scatter of $\pm 0.85h^{-1}$Mpc.  Evidently, even though the reconstructed $\xi_{\rm Lag}$ is not as close to $\xi_{\rm L}$ as in Refs.\cite{LPlaguerre} and \cite{HODlaguerre} -- presumably because the smearing scale at $z=0$ is so large -- Laguerre reconstruction improves the accuracy {\em and} precision of the distance scale estimate.\footnote{Figure~6 in \cite{HODlaguerre} suggests that ignoring the mode-coupling term as we have done here leads to a slight underestimate of $r_{\rm LP}$, so this may be why our reconstructed value is still biased slightly low.}

The other histograms in Figure~\ref{fig:hodrLP} show that, for the other effective volumes as well, the reconstructed values are centered on $\sim 92.1h^{-1}$ with an rms scatter that is slightly smaller for the larger volumes:  they too show an increase in both accuracy and precision compared to $r_{\rm LP}$ in the original $\xi_0$ (i.e. prior to Laguerre reconstruction).

It is interesting to contrast this with the accuracy and precision which result from fixing $\Sigma$ to a fiducial value (in this case, $10.34h^{-1}$Mpc) and only determining the $a_k$ from the fits.  Doing so does not change the mean $r_{\rm LP}$, but the error bar is about 30\% smaller ($\pm 0.6h^{-1}$Mpc rather than $\pm 0.85h^{-1}$Mpc).  This is not surprising:  as Ref.\cite{LPlaguerre} discuss, fixing $\Sigma$ to a fiducial value in this way is a little like performing the reconstruction step with a prior on the background cosmological model, and this artificially reduces the estimated error bar.  In effect, determining $\Sigma$ (in addition to the $a_k$) from the fit and using the best-fit $\Sigma$ to reconstruct frees one from this dependence on a fiducial model.  In this sense, the estimate of $r_{\rm LP}$ which results (the one shown in Figure~\ref{fig:hodrLP}) has been marginalized over the a priori unknown value of $\Sigma$.  

Evidently, even in this extreme smearing scenario, the Laguerre reconstruction methodology -- which makes no assumption about the expected shape of the BAO signal -- returns a distance scale estimate that is accurate to sub-percent precision for volumes that are larger than $\sim 10h^{-3}$Gpc$^3$.  Since future surveys target similar comoving volumes but at higher redshifts where the smearing is smaller, we expect our methodology to return sub-percent precision on the estimated distance scale.  Of course, for smaller survey volumes, other datasets may provide better constraints on $\Sigma$, and hence on the prior distribution one should use when marginalizing.  We discuss how one might proceed in such cases in Section~\ref{sec:hermites}.

\subsection{Other uses of the estimated smearing scale}\label{sec:other}
Our Laguerre-based estimate of the smearing scale is particularly interesting as $\Sigma$ potentially provides an estimate of the amplitude of $P_{\rm L}(k,z)$, and hence the linear theory growth factor, that is not degenerate with the bias of the tracers (but see Ref.\cite{rsdPeaks} for why this may not be exactly true).  Crudely speaking, this is because on the scales which dominate the integrand in equation~(\ref{smear}), the power spectrum has approximately the same shape as $P_{\rm L}$, only its amplitude is different:  $P_{\rm obs}(k)\approx b^2\,P_{\rm L}(k)$.  Therefore, if one defines $\Sigma_{\rm obs}^2$ by inserting $P_{\rm obs}$ in place of $P_{\rm L}$ in equation~(\ref{smear}), then the ratio $\Sigma_{\rm obs}/\Sigma\approx b$.  

In practice, we must apply this methodology to the monopole of the redshift space clustering signal, for which
\begin{equation}
  \frac{\Sigma_{\rm obs}}{\Sigma_{\rm eff}} \approx b_{\rm eff}
\left[1 + \frac{f(2+f)}{3}\frac{1 + 6\beta/5 + 3\beta^2/7}{1 + 2\beta/3 + \beta^2/5} \right]^{-1/2}
  \label{betaf}
\end{equation}
depends on both $f$ and $\beta$ rather than $b$ alone.  If $\beta$ is determined from the angular dependence of the clustering signal (e.g. the ratio of the monopole to the quadrupole) then this can be combined with equation~(\ref{betaf}) to estimate $f$.  
We intend to explore this estimate of $f$ in future work.

\section{The Hermite limit}\label{sec:hermites}
The previous section noted that it may be interesting to search for other parameterizations of $\xi_{\rm L}$ which return tighter constraints on the smearing scale.  However, if the tracers are sufficiently sparse that the uncertainties on $\Sigma$ become large, then the constraint on $\Sigma$ may not be sharp enough to be interesting (either for constraining $b$ or for reconstructing the shape of $\xi_{\rm L}$).  If one must use constraints on the smearing scale from other datasets to perform BAO reconstruction, then it is interesting to ask if alternative parametrizations (to Laguerre) simplify the process of marginalizing over the value of the smearing scale when reconstructing $\xi_{\rm L}$.  As the Introduction notes, this provides additional motivation for exploring other parametrizations of $\xi_{\rm L}$.

To address this, we begin with equation~(\ref{xiNLconv}), and consider the limit in which $r\gg\Sigma$ and $s\gg\Sigma$ (Figures~\ref{fig:pdfSmear} and~\ref{fig:lagSmear} show that the smearing scale is indeed much smaller than the BAO scales of interest).  Then $2\,\sinh(rs/\Sigma^2)\approx {\rm e}^{rs/\Sigma}$
making 
\begin{equation}
  \xi_{\rm NL}(s)\approx \int_0^\infty \frac{dr\,r^2}{\Sigma^3}\,\frac{{\rm e}^{-(r-s)^2/(2\Sigma^2)}}{\sqrt{2\pi}\,rs/\Sigma^2}\,\xi_{\rm L}(r) ,
  \label{xiNL}
\end{equation}
so 
\begin{equation}
  s\,\xi_{\rm NL}(s)\approx \int_{-\infty}^\infty dr\,
  \frac{{\rm e}^{-(r-s)^2/(2\Sigma^2)}}{\sqrt{2\pi}\,\Sigma}\,r\,\xi_{\rm L}(r).
  \label{sxiNLs}
\end{equation}
Note that we can extend the lower limit of integration down to $-\infty$ only if $r-s\ll \Sigma$.  If we parametrize $\xi_{\rm L}$ using a simple polynomial (i.e. equation~\ref{xiLin}), then the integral above can be done analytically.  It is easy to check that each $a_k$ multiplies $(\Sigma/{\cal R})^k$ times a polynomial in $s/\Sigma$.  This polynomial is the same as that which appears in the Laguerre expansion $\mu_k(s/\Sigma)$, when one takes the $s\gg \Sigma$ limit ($E_1\to 1$ and $E_2\to 0$ in equation~A3 of Ref.\cite{LPlaguerre}).  

There is no a priori reason for parametrizing  $\xi_{\rm L}$ with a simple polynomial.  If we parametrize $r\xi_{\rm L}(r)$ using the probabilist's Hermite polynomials instead, 
\begin{equation}
  r\xi_{\rm L}(r) = \sum_{k=0}^n a_k\,H_k\left(\frac{r-r_{\rm fid}}{{\cal R}}\right) ,
  \label{xiHermite}
\end{equation}
then Appendix~\ref{sec:modHermites} shows that the integral in equation~(\ref{sxiNLs}) can still be done analytically.  For each $k$, the result is $(s-r_{\rm fid})^k/{\cal R}^k$ plus additional terms which are lower order polynomials in $s$ multiplied by terms proportional to $({\cal R}/\Sigma)^2$.  By carefully grouping these other terms it is possible to find that $r\xi_{\rm L}(r)$ which, when inserted in equation~(\ref{sxiNLs}), produces a simple polynomial in $s$. Namely, if 
\begin{equation}
  r\xi_{\rm L}(r) = \sum_{k=0}^n a_k\,{\cal H}_k\left(\frac{r-r_{\rm fid}}{{\cal R}},\frac{\Sigma}{{\cal R}}\right) ,
  \label{modHermite}
\end{equation}
where the ${\cal H}_k$ are given in Appendix~\ref{sec:modHermites}, then
\begin{equation}
  s\xi_{\rm NL}(s) =  \sum_{k=0}^n a_k\,\left(\frac{s-r_{\rm fid}}{{\cal R}}\right)^k .
  \label{xiSimple}
\end{equation}
Therefore, if we determine the $a_k$ by fitting the simple polynomial of equation~(\ref{xiSimple}) to the observed $s\,\xi_{\rm NL}(s)$, then the reconstructed/deconvolved $r\,\xi_{\rm L}(r)$ is given by equation~(\ref{modHermite}), provided we first assume a value for $\Sigma/{\cal R}$ (which we discuss shortly).  Note that whereas `Laguerre reconstruction' has a simple polynomial as the reconstructed shape of $\xi_{\rm L}(r)$, this `Hermite reconstruction' of $r\xi_{\rm L}(r)$ has a simple polynomial as the nonlinear shape of $s\xi_{\rm NL}(s)$.  

\subsection{Dependence on smearing scale}
In practice, we will not know the correct value of $\Sigma$, so it is interesting to study the sensitivity of Hermite reconstruction to incorrect choices of $\Sigma$.  In this regard, the structure of this Hermite reconstruction problem has two surprising consequences.  First, because $\Sigma$ does not appear in equation~(\ref{xiSimple}), fitting it to the measurements yields {\em no} information about $\Sigma$.  I.e., if we fit $s\xi_{\rm NL}(s)$ to a simple polynomial, then we cannot make a plot like Figure~\ref{fig:lagSmear}!  Second, for the same fitted $a_k$, varying $\Sigma$ only changes the reconstructed shape (equation~\ref{modHermite}).  In contrast, for Laguerre reconstruction, each choice of $\Sigma$ requires a new fit ($\Sigma$ appears in equation~\ref{xiLag} for $\xi_{\rm NL}$).  In this sense, Hermite reconstruction is more efficient than Laguerre -- one only need determine the $a_k$ once.  

Although $s\xi_{\rm NL}(s)$ does not depend on $\Sigma$, the Hermite reconstruction of $\xi_{\rm L}$, which we will refer to as $\xi_{\rm Her}$, does.  So we now turn to the value of $\Sigma$.  There are two natural choices.  One is to treat the Laguerre-based analysis (e.g. Figure~\ref{fig:lagSmear}) as providing a prior on the value of $\Sigma$.  But if this is rather broad (e.g. for sparse tracers), it may be that an alternative approach is more constraining.  Following \cite{HODlaguerre}, this second approach exploits the fact that, for all biased tracers, the smearing scale is expected to be well approximated by equation~(\ref{smear}).  Suppose we use $\Sigma^2_{\rm obs}$ to denote the result of inserting the observed $P_{\rm obs}(k)$ in equation~(\ref{smear}).  On the large scales (small $k$) which dominate the integral (in $\Lambda$CDM models), $P_{\rm obs}(k)\approx b_{10}^2\,P_{\rm L}(k)$, making $\Sigma\approx \Sigma_{\rm obs}/b_{10}$.  In this approximation, the uncertainty in what to use for $\Sigma$ boils down to what to use for $b_{10}$.  

We will use $b_{\rm fid}$ to denote our best guess for this value, and so we define $\Sigma_{\rm fid}\equiv \Sigma_{\rm obs}/b_{\rm fid}$.  This suggests that if we fit the observed correlation function to equation~(\ref{xiSimple}), with ${\cal R}=\Sigma_{\rm fid}$, then the effect of varying $b$ from its fiducial value yields reconstructed 
\begin{equation}
  r\,\xi_{\rm Her}(r) = \sum_{k=0}^n a_k\,
  {\cal H}_k\left(\frac{r-r_{\rm fid}}{\Sigma_{\rm fid}},\frac{b_{\rm fid}}{b_{10}}\right) .   
  \label{xiHer}
\end{equation}
Note that when $b_{10}=b_{\rm fid}$ then ${\cal H}_k \to H_k$:  the reconstructed shape is a simple sum of Hermite polynomials.  (Of course, this is only true if $\Sigma\approx \Sigma_{\rm obs}/b_{10}$.)  This allows a straightforward estimate of the uncertainty on the reconstructed $\xi_{\rm Her}$:  fitting to $\xi_{\rm NL}$ yields the covariance matrix of the fitted $a_k$.  For a given choice of $b_{\rm fid}/b_{10}$, this can be used to produce uncertainty bands around the shape given by equation~(\ref{xiHer}), and further marginalizing over the value of $b_{\rm fid}/b_{10}$ gives the full uncertainty on the reconstructed shape (see \cite{LPlaguerre} for a detailed discussion).  

We must also decide what to use for $r_{\rm fid}$.  We are particularly interested in BAO scales, where the correlation function exhibits a peak and a dip.  Therefore a reasonable way to determine $r_{\rm fid}$ is as follows.  Initially choose $r_{\rm fid}$ arbitrarily -- a reasonable choice would use the expected value of $r_{\rm LP}$ of equation~(\ref{eq:rLP}) in the current best fitting cosmological model.  Then fit equation~(\ref{xiSimple}), with ${\cal R}=\Sigma_{\rm fid}$, to the observed $s\xi_{\rm NL}(s)$.  Find those scales $r_{\rm peak}$ and $r_{\rm dip}$ where $d\xi_{\rm NL}/ds = 0$.  In practice, since
 $d[s\xi]/d\ln s = s\xi + s\, d\xi/d\ln s$, the peak and dip scales in the evolved correlation function are at those $s$ where 
\begin{equation}
  \sum_{k=0}^n a_k\,\left(\frac{s-r_{\rm fid}}{\Sigma_{\rm fid}}\right)^{k-1}
   \left[\frac{ks - (s-r_{\rm fid})}{\Sigma_{\rm fid}}\right] = 0.
\end{equation}
Now set $r_{\rm fid}$ equal to $(s_{\rm peak} + s_{\rm dip})/2$ and re-fit.  Doing this ensures that the higher order polynomials contribute less and less in the vicinity of $r_{\rm LP}$ (see Figure~10 in \cite{LPlaguerre} for an explicit demonstration).  The $a_k$ which result can then be inserted in equation~(\ref{xiHer}).  The peak and dip scales in the reconstruction are where $d[r\xi_{\rm Her}]/d\ln r = r\xi_{\rm Her}$:
\begin{align}
 \sum_{k=0}^n a_k\, \Bigg[&\frac{kr}{\Sigma_{\rm fid}}
    {\cal H}_{k-1} (\frac{r-r_{\rm fid}}{\Sigma_{\rm fid}},\frac{b_{\rm fid}}{b_{10}}) \nonumber\\
    &\quad - {\cal H}_k(\frac{r-r_{\rm fid}}{\Sigma_{\rm fid}},\frac{b_{\rm fid}}{b_{10}}) \Bigg] = 0, 
\end{align}
where we have used the fact that $d{\cal H}_j(x)/dx = j {\cal H}_{j-1}(x)$.  This can be used to determine how the value of $r_{\rm LP}$ in the reconstruction depends on $b$.  One can, of course, weight each of these values by a prior on the value of $b$.

\subsection{The Hermite-reconstructed shape}

\begin{figure}
 \centering
 \includegraphics[width=1\hsize]{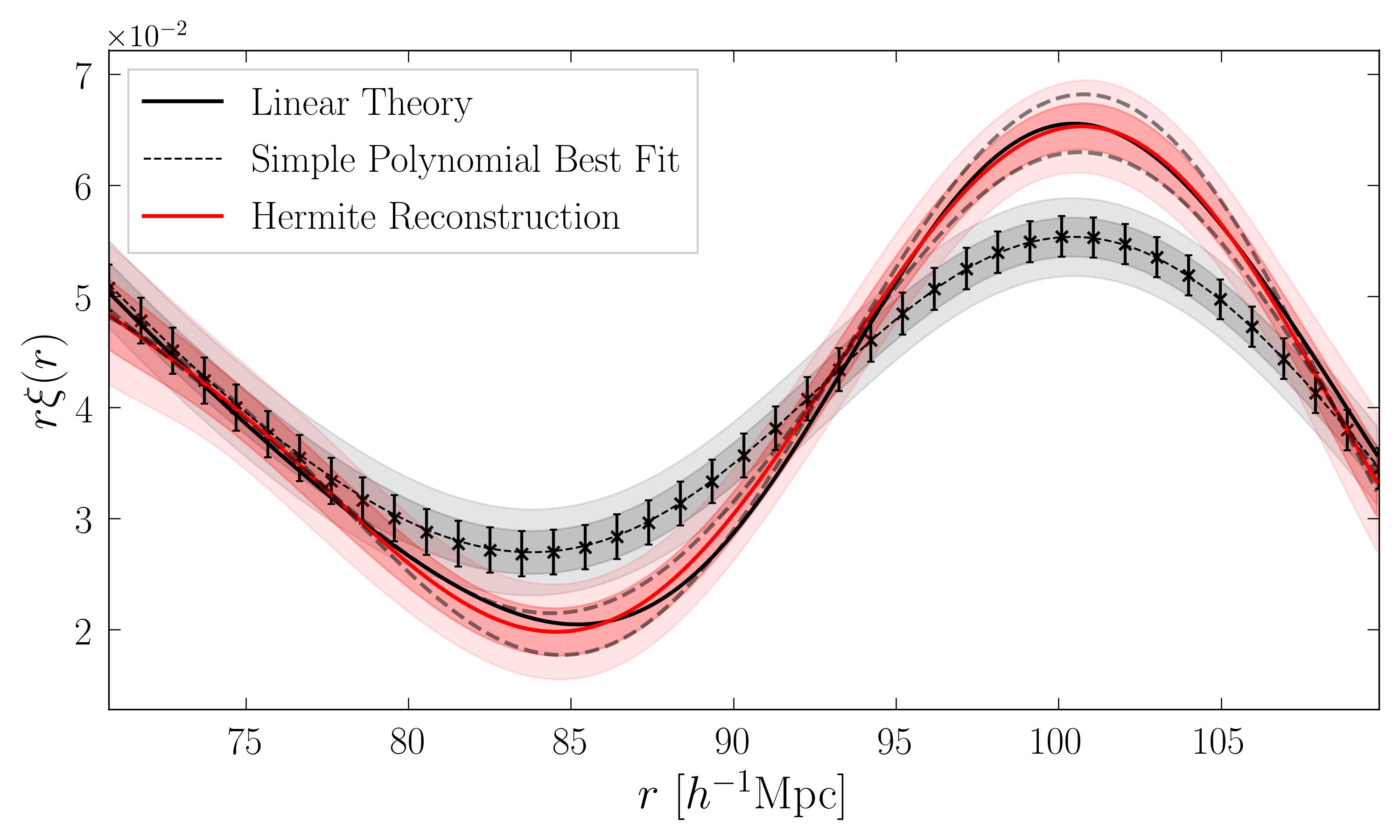}
 \caption{\label{fig:recon}
   Comparison of Hermite reconstruction (equation~\ref{modHermite}) of the shape of the $z=1$ dark matter correlation function (solid red curve; pink bands show 1- and 2-standard deviation uncertainties) with the linear theory shape (solid black) when $\Sigma=5.1h^{-1}$Mpc, the linear theory value, is used.  Dashed grey curves show reconstructions when $\Sigma$ is assumed to be larger or smaller by 10\%.  Symbols with error bars show the measured (i.e. evolved) dark matter correlation function; dashed black curve shows the best fit of equation~(\ref{xiSimple}) to these measurements, which was used to determine the $a_k$ coefficients used in the reconstruction; grey bands show the 1- and 2-standard deviation uncertainties.}
\end{figure}

Figure~\ref{fig:recon} illustrates the various steps associated with Hermite reconstruction.  The symbols with error bars show the measured dark matter correlation function at $z=1$ in an effective volume of $50h^{-3}$Gpc$^3$.  The dashed black curve shows the best fit of equation~(\ref{xiSimple}) with $n=9$, (i.e., a ninth order simple polynomial), to these measurements; associated grey bands are the 1- and 2-standard deviation uncertainties.  This best fit determines the coefficients $a_k$.  The solid red curve with pink error bands shows the associated Hermite reconstruction (equation~\ref{modHermite} with the $a_k$ determined from the fit to the symbols, and $\Sigma=5.1h^{-1}$Mpc or, equivalently, equation~\ref{xiHer} with $b_{\rm fid}=b_{10}=1$) and corresponding uncertainties.  This red curve should be compared with the solid black one, which shows linear theory.  Evidently, when the correct smearing scale is assumed, then Hermite reconstruction works quite well.  (It appears to work much better than the Laguerre reconstructions shown in Figure~\ref{fig:hodrec} only because here we are working with the dark matter at $z=1$ rather than redshift space distorted mock galaxies at $z=0$.  In fact, for the $z=1$ dark matter, the reconstructed shapes, $\xi_{\rm Lag}$ and $\xi_{\rm Her}$, are very similar.)

Dashed grey curves show the Hermite reconstructions when the smearing scale is assumed to be larger or smaller by 10\% (the $a_k$ are the same, of course).  Comparison with Figure~4 in Ref.~\cite{LPlaguerre} shows that, in all cases, the linear theory shape is recovered at least as well as it is for Laguerre reconstruction.  We have also checked that $r_{\rm LP}$ of equation~(\ref{eq:rLP}) in the Hermite reconstructions depends on smearing scale similarly to the Laguerre reconstructions:  weakly (see, e.g., Figure~6 of Ref.~\cite{LPlaguerre}).  Therefore,  reconstructed distance scale estimates and their uncertainties from the Hermite reconstructions are comparable to those from Laguerre reconstruction. This is reassuring because Laguerre reconstruction of the distance scale is accurate, precise and fast \cite{LPlaguerre,HODlaguerre}.  Moreover, as we noted above, Hermite reconstruction is even more efficient, requiring only a single determination of the coefficients $a_k$.

\subsection{Discussion}\label{sec:disc}
One might have thought that how one chooses to parametrize the nonlinear correlation function is of little consequence -- provided the goodness of fit is acceptable.  Our analysis has shown that some parametrizations are more useful than others.  The Laguerre parametrization of $\xi_{\rm NL}$ (equation~\ref{xiLag}) has smearing scale $\Sigma$ dependence in $\xi_{\rm NL}$ but none in $\xi_{\rm L}$, whereas our modified-Hermites (equation~\ref{xiHermite}) have no $\Sigma$ dependence in $s\xi_{\rm NL}(s)$ but some in $r\xi_{\rm L}(r)$. As a result, the Laguerre-based parametrization of $\xi_{\rm NL}$ constrains $\Sigma$ (Figure~\ref{fig:lagSmear}) whereas the simple polynomial parametrization of $s\xi_{\rm NL}(s)$ associated with equation~(\ref{xiHermite}) does not.  (Some of this is a consequence of ignoring the mode-coupling term.  Had we included it, then the Laguerre approach would have no additional $\Sigma$ dependence in $\xi_{\rm NL}$ whereas the modified-Hermites approach would.  However, in practice, the mode-coupling contribution is too small to matter.)

This raises the question of whether or not there is a parametrization of $\xi_{\rm NL}$ which yields tighter constraints on $\Sigma$.  While one is allowed to fit data with a model whether or not the model makes good physical sense, our results suggest that a parametrization which is closer to the physics will fare better:  the modified-Hermites have the shape of the linear theory correlation function depending on time (because $\Sigma$ depends on time), which is unphysical. In contrast, the Laguerre parametrization is consistent with the physics.

While a detailed investigation of physically reasonable parametrizations is beyond the scope of this work, we have performed the following test.  We parametrize $\xi_{\rm L}$ using equation~(\ref{xiHermite}), for which the associated $\xi_{\rm NL}$ is analytic and depends on ${\cal R}/\Sigma$.  This is similar to the Laguerre case, for which $\xi_{\rm L}$ is a simple polynomial which depends on ${\cal R}$ but not $\Sigma$ -- so it is as physically reasonable.  Because of this similarity, we can study how this parametrization constrains $\Sigma$.  We have found that the analogue of Figure~\ref{fig:lagSmear} is almost identical:  there is no significant difference between parameterizing $\xi_{\rm L}(r)$ using simple polynomials or $r\xi_{\rm L}(r)$ using Hermites.  Of course, this does not exclude the possibility that there are other parametrizations which will better constrain $\Sigma$.

\section{Conclusions}
To a good approximation on BAO scales, the evolved correlation function $\xi_{\rm NL}$ is related to the initial one, $\xi_{\rm L}$, by a convolution (equation~\ref{xiNLconv}).  Fitting a series of half-integer Laguerre polynomials (equation~\ref{xiLag}) to the evolved two-point correlation function allows one to constrain the smearing scale $\Sigma$ of the convolution kernel (Figure~\ref{fig:lagSmear}) even when neither the amplitude nor shape of $\xi_{\rm L}$ are known.  In addition, when applied to $\xi_{\rm NL}$ at different redshifts, the method correctly returns the fact that the smearing scale is larger at later times (compare Figures~\ref{fig:lagSmear} and~\ref{fig:hodZmear}).  Our Laguerre approach shows that to constrain the value of $\Sigma$, it is enough to endow the parametrization of the $\xi_{\rm NL}-\xi_{\rm L}$ relation with the correct structure (i.e. one that reflects the fact that the two are related by a convolution).  

In configuration space, the smearing is expected to be approximately independent of the nature of the observed tracers -- i.e. of halo or galaxy bias.  However, in redshift space, the effective smearing is expected to depend weakly on bias (equation~\ref{zmear}); our Laguerre-based estimates of $\Sigma_{\rm eff}$ in redshift-space distorted mock galaxy catalogs are consistent with this expectation (Figure~\ref{fig:hodZmear}).  

In the Laguerre framework, knowledge of the smearing scale allows one to deconvolve and hence reconstruct the shape of $\xi_{\rm L}$ from measurements of $\xi_{\rm NL}$, without any prior assumptions about the shape or amplitude of $\xi_{\rm L}$.  As the shape and amplitude of the reconstructed correlation function can be used to constrain cosmological parameters, the Laguerre methodology can be used to provide more realistic estimates of the precision of the constraints.  In particular, the estimated accuracy and precision of the Laguerre-reconstructed contraints do not depend on choosing a fiducial cosmological model.  We demonstrated this for the linear point feature (equation~\ref{eq:rLP}) in the reconstructed $\xi_{\rm L}$ (Figures~\ref{fig:hodrec} and~\ref{fig:hodrLP}).

In practice, such constraints will depend on the nature of the biased tracers and the volume of the survey.  For small survey volumes, the constraint on the smearing scale is not tight, so marginalizing over its value can significantly weaken constraints on cosmological parameters.  When this occurs, it may be preferable to use tighter constraints on $\Sigma$ which come from other datasets, and then, we argued that the full Laguerre-based analysis may not be necessary.  Provided that $\Sigma$ is a small fraction of the scales of interest, the simple polynomial  - modified Hermites (equation~\ref{xiHer} with Appendix~\ref{sec:modHermites}) combination for $s\xi_{\rm NL}(s)$ and $r\xi_{\rm L}(r)$ provides a more efficient way of marginalizing over the value of $\Sigma$ when quantifying the accuracy and precision on the distance scale estimate (Figure~\ref{fig:recon} and associated discussion).

We illustrated many of our points using mock galaxy catalogs at $z=0$, where the smearing is so large that the BAO feature in $\xi_{\rm NL}$ is nearly completely smeared out.  As our Laguerre-based results were promising nevertheless, we are in the process of implementing the ideas presented here in realistic mock galaxy catalogs which are more relevant to the next generation of cosmological surveys.



Finally, although we have focussed on the BAO smearing scale, recent work has highlighted the benefits of combining full-shape analyses of galaxy power spectra with BAO distance scale estimates to constrain cosmological parameters \cite{fullshapePk}.  Since both Hermite and Laguerre reconstructions reproduce the full shape of $\xi_{\rm L}$ over a rather broad range of scales, their speed and simplicity enable the development of a similar program in configuration rather than Fourier space.  We hope this feature of our reconstructions is exploited in future work.


\begin{acknowledgments} 
  FN and RKS thank the Munich Institute for Astro- and Particle Physics (MIAPP) which is funded by the Deutsche Forschungsgemeinschaft (DFG, German Research Foundation) under Germany's Excellence Strategy – EXC-2094 – 390783311, for its hospitality during the summer of 2019.
  RKS is grateful to the ICTP and the IFPU in Trieste for their hospitality during the summer of 2021.
  FN acknowledges support from the National Science Foundation Graduate Research Fellowship (NSF GRFP) under Grant No. DGE-1845298.
  IZ acknowledges support from NSF grant AST-1612085.
\end{acknowledgments}

\bibliography{hermiteLaguerre}

\appendix

\section{Dependence of $\chi^2$ on effective volume}\label{sec:chi2}
Sufficiently close to the best fit (the minimum of the $\chi^2$), we expect the $\chi^2$ curve to be approximated by a quadratic function. We can expand it around the best fit coefficients $\hat{\textbf{a}}$ as
\begin{equation}
\chi^2(\textbf{a}) = \chi^2(\hat{\textbf{a}}) + {\cal D} \cdot (\textbf{a} - \hat{\textbf{a}}) 	+ \frac{1}{2} (\textbf{a} - \hat{\textbf{a}})^T \cdot {\cal M} \cdot (\textbf{a} - \hat{\textbf{a}}),
\end{equation}
where ${\cal D}_i = \partial \chi^2/\partial a_i$ is the first derivative vector and ${\cal M}_{ij} = \partial^2 \chi^2/\partial a_i \partial a_j$ is the Hessian matrix. The best fit coefficients $\hat{\textbf{a}}$ will be determined from the fact that the first order derivatives $\partial \chi^2/\partial a_i$ are zero for all $a_i$ at the minimum $\chi^2_{\rm min}= \chi^2(\hat{\textbf{a}})$, so we can write the above equation as
\begin{equation}
	\Delta\chi^2\equiv\chi^2(\textbf{a}) - \chi^2_{\rm min} = (\textbf{a} - \hat{\textbf{a}})^T \cdot  ({\cal M}/2) \cdot (\textbf{a} - \hat{\textbf{a}}).
\end{equation}

The $\chi^2$ function can be expressed as the following weighted inner product
\begin{equation}\label{chi2_def}
	\chi^2 = (\xi - \hat{\xi})^T \cdot {\cal C}^{-1} \cdot (\xi - \hat{\xi}),
\end{equation}
where ${\cal C}^{-1}$ is the inverse of the covariance matrix, $\xi$ is the measurement, and $\hat{\xi}$ is our best fit. As we discussed in the text, if the value of the smearing scale $\Sigma$ is known, then our fitting model is linear in coefficients. For a linear model, the best fit can be written in matrix form as $\hat{\xi} = {\cal A} \cdot \hat{\textbf{a}}$, where ${\cal A}$ is the design matrix. In our case, the elements of the design matrix are ${\cal A}_{ij} = {\cal H}_j\left([r_i-r_{\rm fid}]/\Sigma_{\rm fid}, b_{\rm fid}/b_{10}\right)$. Inserting this linear model in equation~(\ref{chi2_def}) and calculating the associated Hessian matrix yields 
\begin{equation}
  \Delta\chi^2 = 
  (\textbf{a} - \hat{\textbf{a}})^T \cdot ({\cal A}^T {\cal C}^{-1} {\cal A}) \cdot (\textbf{a} - \hat{\textbf{a}}).
 \label{chi2_lin}
\end{equation}
The covariance matrix ${\cal C}$ scales as the inverse of the effective survey volume, so ${\cal C}^{-1}$ and consequently $\chi^2(\textbf{a}) - \chi^2_{\rm min}$ scale as the effective volume. 

Let $\delta \textbf{a} = \textbf{a} - \hat{\textbf{a}}$ be a change in the fitting coefficients whose first element is arbitrary $\delta a_1$, but the rest of whose elements are selected to minimize the $\Delta\chi^2 = \chi^2(\textbf{a}) - \chi^2_{\rm min}$. Then $\delta a_1$, the uncertainty on the value of $a_1$ is
\begin{equation}
	\delta a_1 = \pm \sqrt{\frac{\Delta \chi^2}{{[\cal C}^{-1}]_{11}}},
\end{equation}
which scales as the inverse of the square root of the effective volume.

For the nonlinear case (when $\Sigma$ is not known) equation~(\ref{chi2_def}) is still valid. By calculating the second derivatives in the general form, $\Delta \chi^2$ of the nonlinear model can be written as
\begin{equation}
	\Delta\chi^2 = \delta \textbf{a}_{k} \, \Big[\frac{\partial \hat{\xi}_{i}^T}{\partial a_k} [{\cal C}^{-1}]_{ij}  \frac{\partial \hat{\xi}_j}{\partial a_l} - (\xi - \hat{\xi})_{i}^T [{\cal C}^{-1}]_{ij} \frac{\partial^2 \hat{\xi}_j}{\partial a_k \partial a_l} \Big]_{kl} \, \delta \textbf{a}_l,
 \label{chi2nl}
\end{equation}
where all the derivatives are evaluated at the best fit values of the fitting parameters.  

In the linear model, the second derivatives are all zero, and this expression reduces to equation~(\ref{chi2_lin}).  In a nonlinear model, finding the best fit $\hat{\xi}$ must proceed iteratively, and then we only need to insert the solution into equation~(\ref{chi2nl}). Since the inverse of the covariance matrix appears in both the first and second terms in the square brackets above, the scaling with effective volume for the nonlinear case is the same as for the linear one.

As noted in the main text, to estimate the best-fitting $\Sigma$, we solve a linear least square problem for a number of choices of $\Sigma$. We then compute the $\chi^2$ values and choose the $\Sigma$ that minimizes the $\chi^2$. For this step and measuring the uncertainty on the best-fitting $\Sigma$ for each realization, we fit a quadratic function $A_0 (\Sigma - A_1)^2$ to the $\chi^2$ values as a function of $\Sigma$. The $A_1$ parameter gives us the best-fitting $\Sigma$, and then by setting the confidence level of $\Delta\chi^2 = 1$, the uncertainty on $\Sigma$ can be determined by $1/\sqrt{A_0}$. 

For the error bar on $r_{\rm LP}$, we need to propagate the uncertainty from the fitted parameters of the correlation function to the position of the peak and the dip, and finally to the linear point. In this nonlinear case that $\Sigma$ is unknown, we should write the linear point position as a function of the polynomial coefficients $\{a_k\}$s and $\Sigma$, and then expand the result around the best-fit parameters. The error bar on $r_{\rm LP}$ can be written as
\begin{equation}
	\sigma_{\mathrm{LP}}=\left\{\sum_{i, j} \frac{\partial r_{\mathrm{LP}}}{\partial \beta_{i}} [\operatorname{Cov}(\overline{\boldsymbol{\beta}})]_{i j} \frac{\partial r_{\mathrm{LP}}}{\partial \beta_{j}}\right\}^{1 / 2},
\end{equation}
where $\beta = \{a_0, a_1, \dots, a_k, \Sigma\}$.

\section{Modified Hermite polynomials}\label{sec:modHermites}
We are interested in integrals of the form 
\begin{equation}
 I_k(y,\beta) \equiv \int dx\,\frac{{\rm e}^{-(y-x)^2/2}}{\sqrt{2\pi}}\,H_k(\beta x),
\end{equation}
where the $H_k(x)$ are the probabilist's Hermite polynomials and $\beta>0$. 

We start with the fact that when $\beta=1$ then 
\begin{equation}
 I_k(y,1) = \int dx\,\frac{{\rm e}^{-(y-x)^2/2}}{\sqrt{2\pi}}\,H_k(x) = y^k.
\end{equation}
We then note that 
\begin{equation}
 H_n(\beta x) = \sum_{i=0}^{[n/2]} \beta^{n-2i} \, (\beta^2 - 1)^i 2^{-i} {n\choose 2i} \frac{(2i)!}{i!} H_{n-2i}(x) .
\end{equation}
This with the definition of $I_k(y,1)$ show that when $\beta\ne 1$, then $I_k(y,\beta)$ is a sum of polynomials which are each multiplied by different powers of $\beta$.  E.g., the term of highest order in $y$ is $(\beta y)^k$.  We can remove all the other terms by subtracting appropriate combinations of $\beta$ and $H_j(\beta x)$.  Doing so defines the functions called ${\cal H}_k(x,\beta)$ in the main text.  Explicitly, they are:
\begin{align}
  {\cal H}_0(\beta x) &= H_0(\beta x) \nonumber\\
  {\cal H}_1(\beta x) &= H_1(\beta x) \nonumber\\
  {\cal H}_2(\beta x) &= H_2(\beta x) - A H_0(\beta x) \nonumber\\
  {\cal H}_3(\beta x) &= H_3(\beta x) - 3A H_1(\beta x) \nonumber\\
  {\cal H}_4(\beta x) &= H_4(\beta x) - 6A\, H_2(\beta x) + 3 A^2\,H_0(\beta x)\\
  {\cal H}_5(\beta x) &= H_5(\beta x) - 10A\, H_3(\beta x) + 15 A^2\,H_1(\beta x)\nonumber\\
  {\cal H}_6(\beta x) &= H_6(\beta x) - 15A\, H_4(\beta x) + 45 A^2\,H_2(\beta x) \nonumber\\ 
  &\quad - 15 A^3\, H_0(\beta x) \nonumber\\
  {\cal H}_7(\beta x) &= H_7(\beta x) - 21A\, H_5(\beta x) + 105 A^2\,H_3(\beta x) \nonumber\\
  &\quad - 105 A^3\, H_1(\beta x)\nonumber\\
  {\cal H}_8(\beta x) &= H_8(\beta x) - 28A\, H_6(\beta x) + 210 A^2\,H_4(\beta x) \nonumber \\
  & \quad - 420 A^3\, H_2(\beta x) + 105A^4\,H_0(\beta x)\nonumber\\
  {\cal H}_9(\beta x) &= H_9(\beta x) - 36A\, H_7(\beta x) + 378 A^2\,H_5(\beta x) \nonumber \\
  & \quad - 1260 A^3\, H_3(\beta x) + 945A^4\,H_1(\beta x)\nonumber
\end{align}
where $A\equiv \beta^2 - 1$.  This shows that ${\cal H}_k \to H_k$ as $\beta\to 1$.  
Moreover, note the similarity of the structure to that for the $H_k$ themselves:
the numerical coefficients are the same as for $H_k$, with $x^j\to H_j$ and with each extra term receiving an additional power of $A$.  This similarity means that $d{\cal H}_j(y)/dy = j{\cal H}_{j-1}(y)$, which mirrors the fact that $dH_j(x)/dx = jH_{j-1}(x)$.

\end{document}